\documentclass[a4paper]{jpconf}
\usepackage[utf8]{inputenc}
\usepackage{graphics}
\usepackage{lscape}
\usepackage{amsmath} 
\usepackage{amsfonts} 
\usepackage{amssymb}
\usepackage{textcomp}
\usepackage{setspace}
\usepackage{multicol} 
\usepackage{tabularx}
\usepackage{array}
\usepackage{enumitem}
\usepackage{eurosym}

\usepackage{booktabs}
\usepackage{longtable}
\usepackage{bbm,fancyhdr,epsfig}
\usepackage{color}
\usepackage{colortbl}
\usepackage{colordvi}
\usepackage{threeparttable}
\usepackage{multirow}    
\usepackage{framed}
\usepackage{booktabs}
\definecolor{dunkelgrau}{rgb}{0.8,0.8,0.8}
\definecolor{hellgrau}{rgb}{0.95,0.95,0.95}
\definecolor{shadecolor}{gray}{.65}
\definecolor{TableHeadGray}{gray}{.8}

\usepackage{graphicx}
\begin{document}

\title{Status of High-Energy Neutrino Astronomy}

\author{Marek Kowalski}

\address{Deutsches Elektronen-Synchrotron (DESY), Platanenallee 6, D-15738, Zeuthen;  Humboldt-Universit\"at zu Berlin, Institut f\"ur Physik, Newtonstrasse 15, D-12589, Berlin}

\ead{marek.kowalski@desy.de}

\begin{abstract}
With the recent discovery of high-energy neutrinos of extra-terrestrial origin by the IceCube neutrino observatory, neutrino-astronomy is entering a new era. This review will cover currently operating open water/ice neutrino telescopes, the latest evidence for a flux of extra-terrestrial neutrinos and current  efforts in the search for steady and transient neutrino point sources. Generalised constraints on potential astrophysical sources are presented, allowing to focus the hunt for the sources of the observed high-energy neutrinos. 

\end{abstract}

\section{Introduction}

 During the last decades,  considerable efforts were made to establish high-energy neutrinos as  cosmic messenger. The motivation derives from the special properties of neutrinos. They interact only weakly and hence can escape energetic and dense astrophysical environments otherwise opaque to electro-magnetic radiation. It is for that reason that neutrinos promise to provide unique insights into a number of extreme astrophysical phenomena, ranging from stellar explosions to particle acceleration in the center of active galaxies. 
Furthermore, the observation of high-energy neutrinos can help to decipher the more than 100 year old mystery of cosmic rays. Charged cosmic rays are expected to produce high-energy neutrinos through the interaction with ambient matter or radiation fields. However, while charged cosmic rays are deflected by magnetic fields on the way from the source to the Earth, neutrinos always point back to their origin, unaffected by matter or radiation fields encountered on their path. Hence high-energy neutrinos can provide direct information about the accelerators producing the highest-energy cosmic rays.  Accordingly, with the recent discovery of high-energy neutrinos of extra-terrestrial origin by the IceCube neutrino observatory 
\cite{science13,hese3yr}, neutrino-astronomy has made a long-awaited step. 

In this contribution I review the arguments why an astrophysical flux of neutrinos was expected to be observed (chapter 2), the currently operating open water/ice neutrino telescopes (chapter 3), the latest evidence for a flux of extra-terrestrial neutrinos and current  efforts in the search for steady and transient neutrino point sources (chapter 4). Finally,  generalised constraints on potential astrophysical sources are being discussed, allowing to focus the hunt for the sources of the observed high-energy neutrinos (chapter 5).

\section{Expectations}
The existence of high-energy astrophysical neutrinos has been  motivated by the observations of high-energy hadronic cosmic rays. At their acceleration sites, but also during propagation through space, cosmic rays produce neutrinos in the collision with ambient matter and radiation fields. The resonant interaction of a high-energy proton ($p$) with a target photon ($\gamma$), 
\begin{equation} 
p+\gamma \rightarrow \Delta^{+} \rightarrow n+\pi^+, 
\label{eq:delta}
\end{equation} 
results in a charged pion, that decays further to produce neutrinos. From the pion decay a flux ratio of $\phi_{\nu_e}:\phi_{\nu_\mu}:\phi_{\nu_\tau} = 1:2:0$ is expected at the source, which during propagation from the source to Earth changes to  $\phi_{\nu_e}:\phi_{\nu_\mu}:\phi_{\nu_\tau} =1:1:1$ due to neutrino oscillations (small deviations are expected in case of strong magnetic fields inside the sources \cite{nu_oscillations}).  
%The target photon can be a hard (UV, X-ray) photon produced near the source itself, or one of many soft photons from other sources such as the cosmic microwave background.

Detailed model predictions exist for various source classes, ranging from Gamma-Ray Bursts to Active Galactic Nuclei (see e.g.\ \cite{Becker:2007sv} for a review of sources). To provide a rough estimate of the neutrino flux that can be expected, it is sufficient to review an early argument by Waxman \& Bahcall  \cite{waxman_00}. One starts by considering the energy injection rate in protons per unit volume, $\dot{\epsilon}$, that can be constrained by the observed flux of cosmic rays on Earth. Assuming that the cosmic-rays above $10^{19}$~eV consist primarily of protons and are of extra galactic origin, one finds  $\dot{\epsilon}=(0.5\pm0.015) \times 10^{44} {\rm ergs~Mpc^{-3} yr^{-1}} $ \cite{waxman_2013}. The essential argument for formulating an upper bound on the neutrino flux --- i.e. the Waxman-Bahcall bound \cite{waxman_00} --- is that within the sources, the observed cosmic ray protons are expected to have undergone the reaction of Eq.\ \ref{eq:delta} at most once. This is because the reaction transforms the proton to a neutron, that being neutral, can not  be confined anymore by the source magnetic fields. The neutron hence escapes the acceleration region before it eventually decays back to a proton (note that the Lorenz boost of a $10^{18}$~eV neutron provides for a decay length of the the size of a typical galaxy). Through Eq.  \ref{eq:delta}, the cosmic ray energy density is kinematically tied to  to the neutrino energy density and hence to the expected neutrino flux of high-energy neutrinos, leading to an upper bound on the flux of astrophysical neutrino of all flavors \cite{waxman_2013}:

\begin{equation} 
E_\nu^{-2}\phi_{\rm WB}\approx 3 \times 10^{-8} {\rm GeV cm^{-2} s^{-1} sr^{-1}}.
\label{eq:wbflux}
\end{equation} 

The bound only weakly depends on the redshift evolution of a given population, such as the observed AGN luminosity function or GRB rate evolution.  A flux of neutrinos at the level of the Waxman \& Bahcall bound is detectable within a few years of a cubic-kilometre sized detector and served as an important benchmark in evaluating the sensitivity of neutrino telescopes.

\section{Neutrino Detectors}
The concept of open water/ice neutrino telescopes has been successfully demonstrated in several places world-wide by now. The detection principle for these detectors is similar for all detectors: Cherenkov-light produced by charged particles --- either background muons produced in air showers above the detector, or particles produced in a neutrino interaction --- is recorded by a three-dimensional array of photomultipler tubes (PMTs) contained in appropriate pressure-resisting glass housings. The arrival time allows reconstructing the direction of the particles, while the total number of photons recorded is used to reconstruct  its energy. Arrival direction, energy and topology allows to distinguish  background from neutrino-induced events. A more refined analysis is then needed to distinguish astrophysical neutrinos from atmospheric neutrinos.  
In the following, an overview of the currently operating neutrino detectors is provided. Future projects are covered in a separate contribution to this conference \cite{spiring_ecrs}.  

\subsection{Mediterranean Projects}
The Mediterranean sea is offering deep sites (up to 4.5 km) with very transparent water. Consequently, several efforts where started to exploit the different sites: the NESTOR project deployed off the coast of the Peloponnesus \cite{nestor}, the NEMO project  \cite{nemo} deployed off the coast of Sicily as well as the ANTARES project \cite{antares} deployed off the Provance cost in France. While technologically all projects have had significant impact on the field, only the ANTARES collaboration has been operating a sizeable detector over serval years. Construction of the ANTARES detector was completed in 2008 and it now  consists of 12 strings, each carrying 25 storeys consisting of three PMTs, each. The instrumented volume of ANTARES comprises 0.01 km$^3$.  

\subsection{Lake Baikal}
In the waters of Lake Baikal, the NT200 detector was installed using the frozen lake surface during the winter as a platform for deployment. NT200 was completed in 1998, consisting of 8 strings with 192 PMTs in total. Three more strings carrying 12 PMTs each where added in 2005, resulting in the NT200+ detector \cite{baikal}. The detector reaches a depth of 1100 m and, similar to ANTARES, is still in operation. 
\subsection{South Pole}

The South Pole is home to the  currently largest operating neutrino detector. The IceCube detector  \cite{icecube}, which was deployed between 2005 and 2010, consists of 86 strings with 5160 PMTs in total. The instrumented volume comprises a cubic-kilometre and it is deployed between 1450 and 2450 m depth. Unlike open-water detectors, the deployment of IceCube strings required melting holes into the antarctic ice. Once the instrumented holes are frozen again, the ice provides a very stable environment to operate a detector.  The deep-ice detector  is complemented by a km$^2$-sized air shower array at the surface.

\section{Recent Results}
Large neutrino detectors have significantly gained in sensitivity during the last few years, allowing to address a range of scientific topics, from neutrino oscillations and  dark matter  to neutrino astronomy. In light of the recent discovery of an astrophysical flux of high-energy neutrinos, the focus of this contribution is on the  observation of astrophysical neutrinos. Furthermore, this review does not aim at giving a comprehensive summary of all results, but rather an overview of some of the recent results   that shape our understanding of the newly discovered signal.

\subsection{The discovery of astrophysical Neutrinos}

\begin{figure}[htb]
	\centering
\includegraphics[scale=0.5]{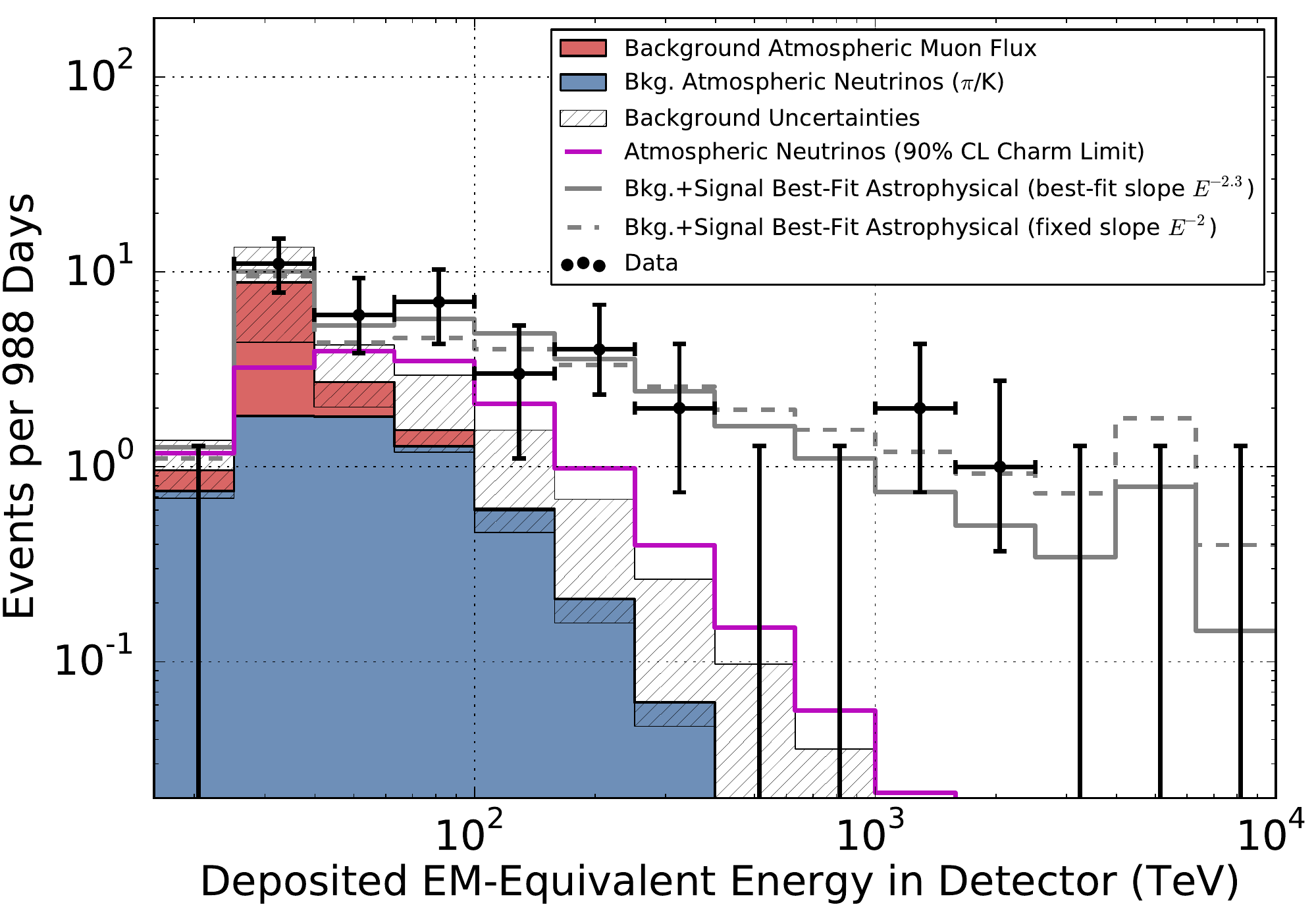}
\caption{The visible deposited energy from the three-year search of starting events in IceCube \cite{hese_2014}.}
\label{fig:hese_spec}
\end{figure}

Evidence for an excess of high-energy neutrinos was seen in data from the incomplete IceCube data from 2008 and 2009, both in the track channel (dominated by $\nu_\mu$) \cite{anne_ic59}, as well as the cascade channel (dominated by $\nu_e$ and $\nu_\tau$) \cite{eike_ic40}. 
The subsequent discovery of two neutrino events above PeV energies in two years of data (between 2010 and 2012) provided more evidence that there is a new unaccounted component in the flux of high-energy neutrinos \cite{Aartsen:2013bka}. %However, the early data would still be consistent with a stronger than expected non-conventional source of atmospheric neutrinos, namely prompt neutrinos from the decay of charmed mesons in the atmosphere. 
None of the independent observations of an excess  by itself quite reached a 3 sigma significance (though combined they would have). More importantly, however, the excess was seen only in the energy spectrum and hence could be explained in principle also by a harder atmospheric neutrino component, such as expected from the decay of charmed mesons in the atmosphere. This changed for a refined analysis of the data, which focused on events starting in the detector  \cite{science13}. By requiring that the neutrino interaction vertex lies within the detector (the outer detector layers provide a veto) and by reconstructing the directions of the events using the full available PMT information, one has a new handle on separating astrophysical from atmospheric neutrinos through a self-vetoing technique \cite{veto}: An atmospheric neutrino would inevitable be produced in coincidence with atmospheric muons from the same air shower. Hence, a downward going neutrino that can be identified without a coincident trough-going (atmospheric) muon track would be rather strong evidence for being of extra-terrestrial origin. The refined analysis using the self-vetoing technique has shown that the majority of the 28  starting neutrino events identified, including the two highest energy events, are consistent with being of extra-terrestrial origin  \cite{science13}. The  analysis was extended from 2 to 3 years of data, resulting in  37 events in total and allowing to reject the atmospheric-only hypothesis with 5.7 $\sigma$  \cite{hese_2014}. The energy spectrum of these events is shown in Fig.\ \ref{fig:hese_spec}.
Assuming a diffuse flux with power-law spectrum and a 1:1:1 flavor ratio, the best fit flux per neutrino flavor is

\begin{equation}
E^2 \phi=1.5\times 10^{-8} (E/100{\rm TeV})^{-0.3} \rm GeV cm^{-2}s^{-1}sr^{-1}.
\label{flux:diffuse}
\end{equation}
The choice to use 100~TeV as the pivot point in the parameterisation minimises the correlation between the spectra index and the normalisation. It is noteworthy that the resulting normalisation coincides with the upper bound predicted by Waxman \& Bahcall. However, because the observed  neutrino energies  are reaching only up to a few PeV, the connection to the  extragalactic cosmic rays observed with 3 to 4 orders of magnitude higher energies is not obvious.

\subsection{Search for steady point sources}
\begin{figure}[htb]
	\centering
\includegraphics[scale=0.7]{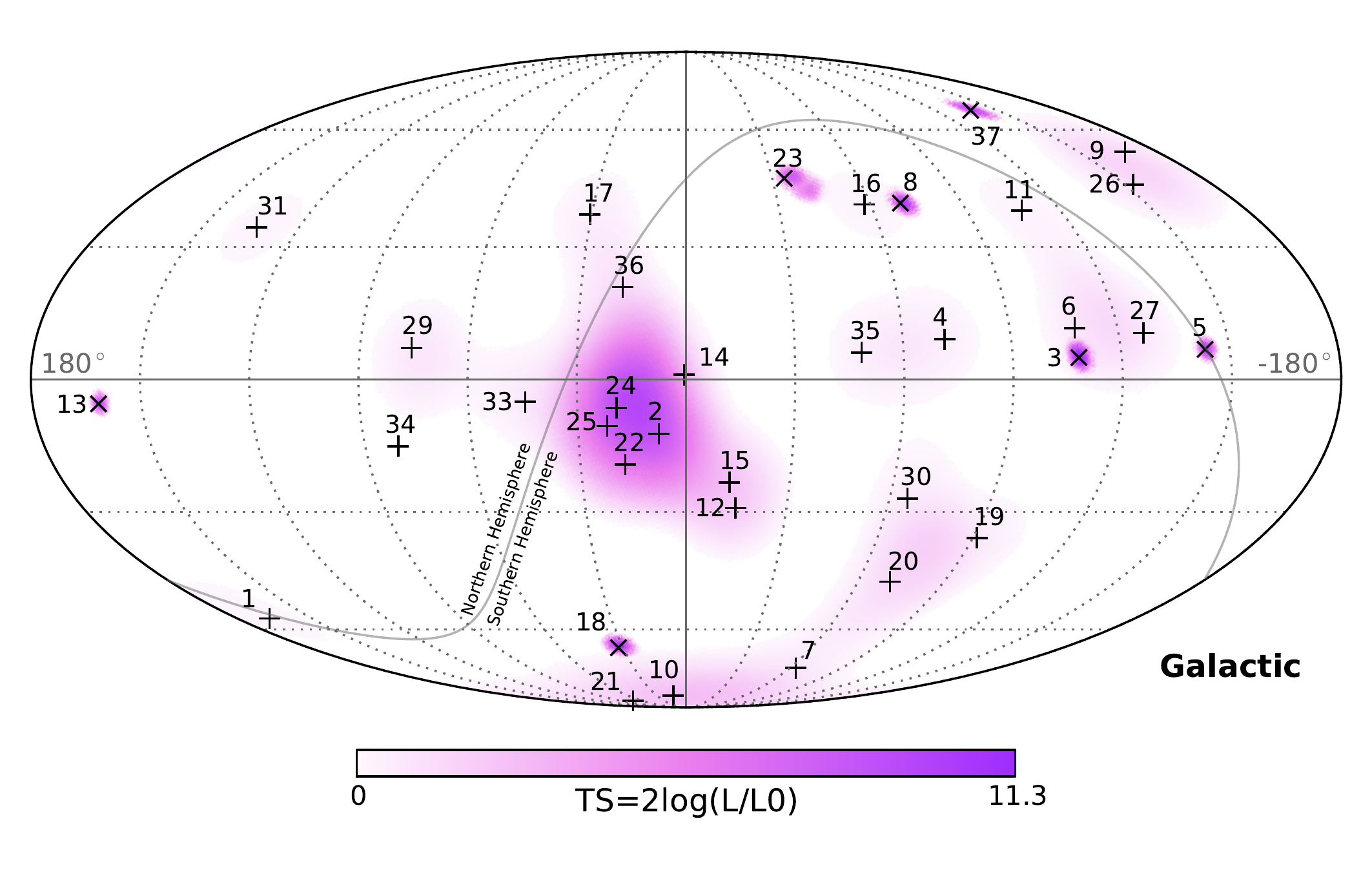}
\caption{Sky map of 37 events from the search for starting events in IceCube \cite{hese_2014}. Most events are cascade-like (marked with $+$) and have an median angular resolution of $15^\circ$, while  the  track-like events (marked with $\times$) are reconstructed with  $\leq 1^\circ$  resolution.  }
\label{fig:skymap}
\end{figure}
The analysis  described in the previous section \cite{science13, hese_2014} produces a sample of  events that is dominated by astrophysical neutrinos. Most events are cascade-like and have an median angular resolution of $15^\circ$, while  the  track-like events are reconstructed with  $\leq 1^\circ$  resolution. The distribution of events in the sky is shown in Fig.\ \ref{fig:skymap}.  So far, the neutrino sky map appears consistent with being isotropic  (if corrected for declination  dependent sensitivity) which is indicative of an extragalactic population of sources. Nevertheless, a sub-dominant galactic component is  possible. The apparent  clustering in Fig.\  \ref{fig:skymap} near the galactic center is the most striking feature of the map. This location in the sky has been studied by the ANTARES collaboration, which because of the favourable field of view of ANTARES can extend the search to lower neutrino energies \cite{Adrian-Martinez:2014wzf}. No excess was observed from the direction of the cluster by ANTARES, which severely constrains the possibility that it is of astrophysical origin.

\begin{figure}[htb]
\centering
\includegraphics[scale=0.7]{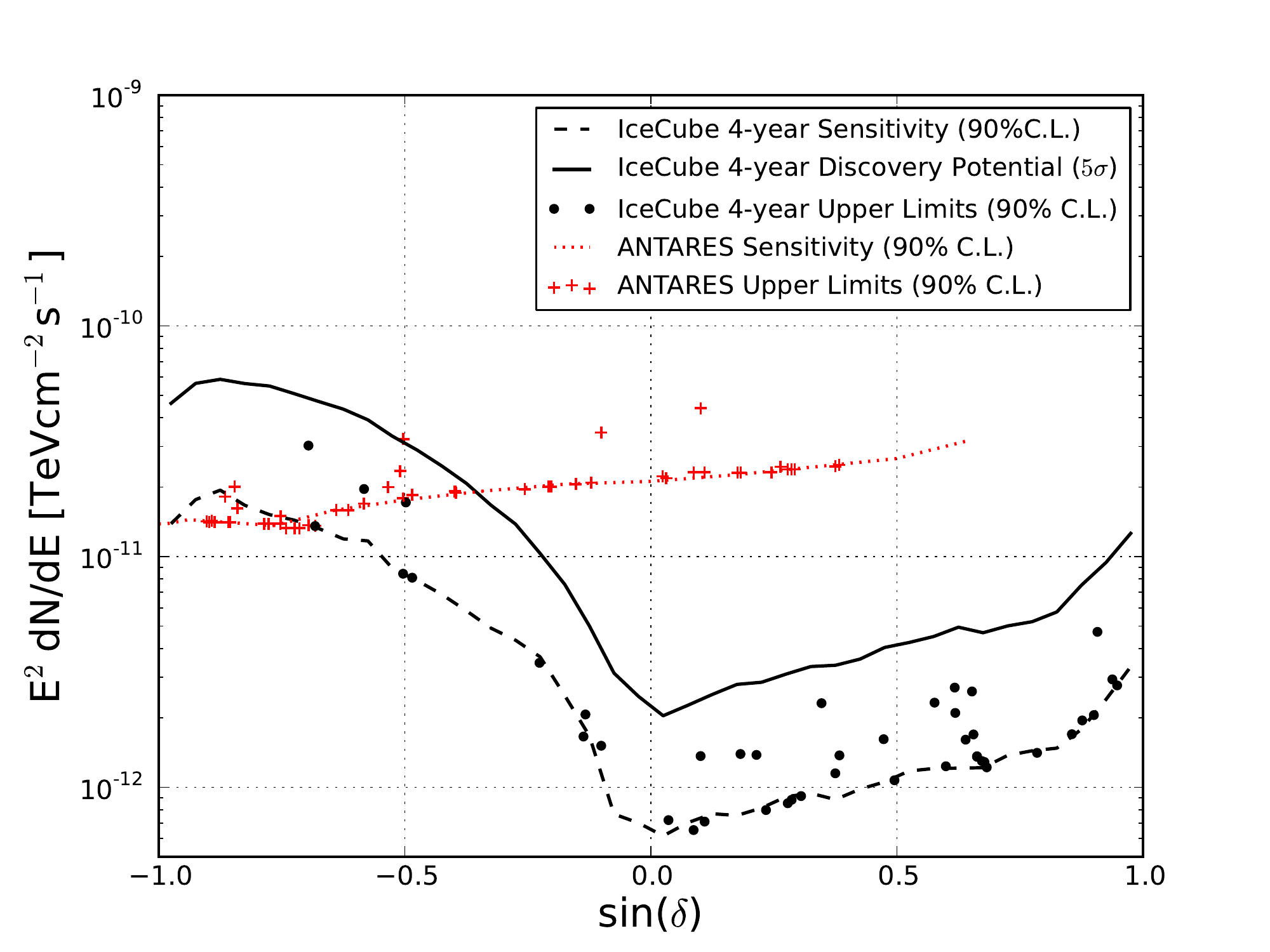}
\caption{Sensitivity to a $E^{-2}$ neutrino flux from individual point sources as a function of the declination \cite{Aartsen:2014cva}. The IceCube constraints are derived from the non-observation of clustering in data dominated by atmospheric muon neutrinos (Northern hemisphere) and atmospheric muons (Southern hemisphere), with correspondingly different energy thresholds \cite{Aartsen:2014cva} (see text for details). The constraints from ANTARES are derived from the non-observation of clustering in a sample dominated by atmospheric neutrinos \cite{Adrian-Martinez:2014hmp}. The symbols represent flux upper limits for individual source candidates.} \label{fig:pslimit}
\end{figure}

The more conventional searches for neutrino point sources use all of the neutrino candidates to search for neutrino clustering or neutrinos associated with catalogs of likely sources. 
So far, no point source was identified by any of the experiments. Figure \ref{fig:pslimit}  shows the latest constraints by IceCube \cite{Aartsen:2014cva} and by ANTARES \cite{Adrian-Martinez:2014hmp}. The IceCube sensitivity was extended to the Southern hemisphere by application of an energy cut to suppress the large background of atmospheric muons. Hence, for the Southern hemisphere the energy threshold of the search is two orders of magnitudes higher than for the search in the Norther hemisphere and the search by ANTARES. (A direct comparison of sensitivities is therefore  model dependent.)  IceCube records about 5 atmospheric neutrino events per year per square-degree northern sky, above which an astrophysical source needs to produce an excess to be detected. Accordingly, the sensitivity to point sources only increases with $\sqrt{T}$ as a function of time, unless further measures are taken to improve the signal-to-noise (such as improving the angular resolution or using energy as an extra discriminator for signal-like events). ANTARES, due to its smaller size,  records about 4 atmospheric neutrinos per day over one hemisphere, hence identification of a point source requires fewer events \cite{Adrian-Martinez:2014hmp}. %Correspondingly the sensitivity grows more linearly with time. 

\subsection{Search for transient sources}
\label{sec:transients}
The most prominent example of a transient source class are Gamma-Ray Bursts (GRB), which  outshine the entire Universe in  $\gamma$-rays during the $\sim1-100$ seconds burst duration. During such a short time the background of atmospheric neutrinos is small ($\ll1$), therefore even a single coincidence of a neutrino with a GRB would be of significant interest. No coincidence with a muon-neutrino event has been reported by any experiment so far. Testing for neutrino emission from 117 GRBs in the Northern hemisphere with IceCube \cite{Abbasi:2012zw} allowed to constrain the aggregate flux from all GRBs to below the Waxman \& Bahcall bound. It also constrains the possibility that GRBs are the sources of cosmic rays \cite{waxman98}, though more data as well as detailed modelling is required to rule-out all scenarios (see e.g.\ \cite{Baerwald:2014zga}).

Other transient source classes could be off-axis or failed GRBs, where the jet does not escape the progenitor star or core collapse SNe with mildly relativistic jets, for which neutrino emission would be expected on time scales similar to GRBs. In such cases, no bright  $\gamma$-ray burst would mark the stellar explosion. To nevertheless obtain sensitivity to a wide range of transient phenomena, neutrino telescopes search for event clusters both in direction and time. These searches are being performed in near realtime, allowing to trigger subsequent follow-up observations. Both ANTARES and IceCube send  target-of-opportunity alerts when neutrino multiplets ---- two or more neutrinos coincident in time and direction --- are observed { \cite{Abbasi:2011ja, Ageron:2011pe}}. (ANTARES also sends alerts for the most  energetic neutrino events). Follow-up observations are performed with the ROTSE and TAROT network of small robotic telescopes, with the Samuel Oschin telescope of PTF, as well as the SWIFT X-ray satellite.

  %%CITATION = ARXIV:1103.4477;%%. 

\section{The larger picture: Generic constraints on possible source classes}

\begin{figure}[htb]
\centering
\includegraphics[scale=0.8]{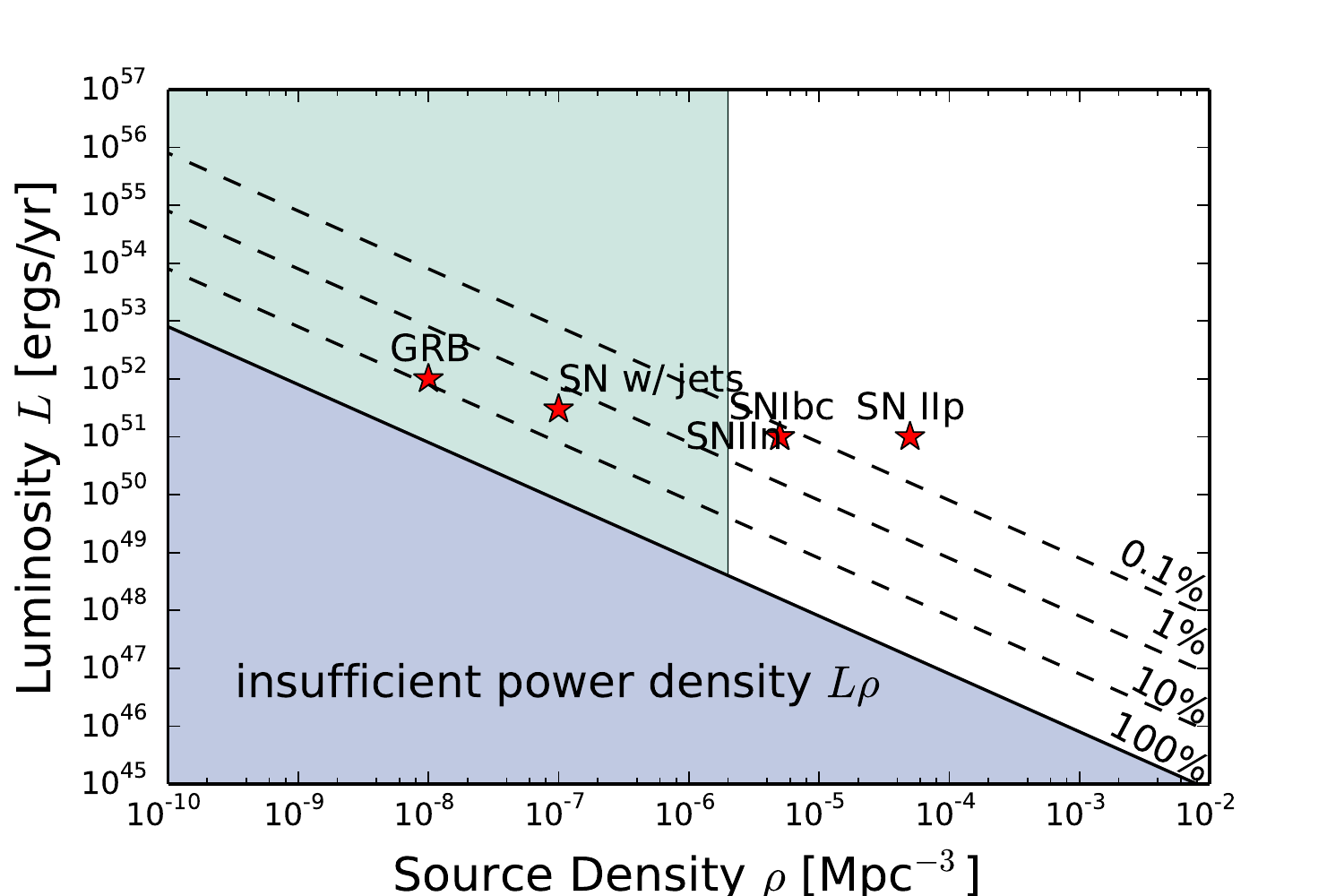}
\caption{
Constraints on potential transient sources responsible for the diffuse astrophysical flux detected by IceCube. The individual source populations are represented by a density$^1$ and typical luminosity$^1$ in a characteristic wavelength band. From left to right: Gamma-Ray Bursts, Supernovae with jets, Supernovae Ib/c, Supernovae IIN and Supernovae IIP. The (full) diagonal line marks the neutrino power density boundary, where the luminosity per volume, $L \cdot \rho $ , saturates the observed diffuse flux; sources above the line release sufficiently abundant energy in electromagnetic radiation to explain the diffuse neutrino flux (dashed lines indicated fractional energy release in neutrinos). Sources in the vertical left (green) region can be excluded as the dominant source of the diffuse flux, since they would result in a signal within dedicated searches of IceCube (see text for details).}
\label{fig1}
\end{figure}

One can use the observation of a diffuse (extragalactic) flux of neutrinos, combined with a non-observation of resolved sources, to constrain the  density of a generic population of sources (see \cite{lipari06} for an early version of the argumentation). In a simplified picture, the bound on the source density can be understood as follows: The product of luminosity per source and source density, $L\cdot \rho$, corresponds to the total emission per volume and is constrained by the observed diffuse flux of neutrinos. For a distance scale to the nearest source $d=(4\pi \rho)^{-1/3}$, along with the power density $L\cdot \rho$, one can estimate the neutrino flux of the nearest object within a population $\phi=L/d^2 \propto \rho^{-1/3}$. The expected flux increases with decreasing source density and the fact that point source searches (including the search for stacked or transient sources\footnote{For transient sources the density refers to the number density of transients in a given year, while the luminosity refers to the energy released by the transient within one year. This choice of terminology allows to generalise the discussion to steady sources.}  have only resulted in upper limits allows us to place constraints on possible populations and their associated source density. In the following, I discuss some current constraints derived from IceCube data using a signal simulation of generic source populations developed by Nora Linn Strothjohann and Andreas Homeier \cite{nora_andreas}.

\subsection{Transient sources}
To start with, the  simulation of a transient population of sources  was assuming number density evolution and luminosity function following that observed for GRBs \cite{wanderman2010}. %Once the fluence is multiplied with the GRB duration, the resulting luminosity function is well represented by a Gaussian distribution with a width of one order of magnitude.
Assuming that the neutrino emission follows the same luminosity function and source population, one can model the neutrino flux expectation. For this, the spectral shape of the individual sources is chosen to match the observed diffuse flux, and the neutrino flux emitted by the total population is normalized to the observed diffuse flux. The simulation allows to predict the number of neutrino muliplets expected to be seen by IceCube (including effective areas and Poisson fluctuations due to low count rates). By requiring consistency with the observed number of neutrino multiplets, i.e. no triplet of neutrinos within 100 seconds \cite{nora_andreas}, one obtains a lower bound on the source density. For a diffuse power-law flux  matching the IceCube flux of Eq.\ \ref{flux:diffuse},
%of $\phi  = 5\cdot10^{-7} \left(\frac{E}{\text{GeV}}\right)^{-2.3} \text{GeV}^{-1} \text{cm}^{-2} \text{s}^{-1} \text{sr}^{-1}$, 
the bound on the source density derived from the non-observation of neutrino triplets or higher multiplicities in three years of IceCube data  corresponds to $\rho\geqslant 2\cdot 10^{-6} \text{Mpc}^{-3}\text{yr}^{-1}$ \cite{nora_andreas}. The bound is broadly consistent with a recent sensitivity estimate \cite{Ahlers:2014ioa} and depends only weakly on the evolution of the GRB density as a function of redshift and on the luminosity function. However, it does depend on the assumed spectral shape, which has been assumed to follow a power-law over the full energy range that IceCube is sensitive to. Fig.~\ref{fig1} illustrates the current constraints on transient sources. The bound from the non-observation of multiplets is shown as a vertical line. As one can see, GRBs, being very rare, are excluded as the dominant sources of the observed diffuse neutrino flux.

On the other hand, core collapse supernovae are still plausible candidates. They possess all the right ingredients for being extraordinary neutrino factories: a) they have been shown to produce ejecta with $10^{50}$ ergs kinetic energy, capable of efficiently accelerating CRs and b) they provide abundant amounts of target material for neutrino production, e.g. the stellar envelope or the circumstellar medium (CSM). The shock acceleration can happen at non-relativistic shock fronts \cite{Murase2010, Katz2011} or within (mildly) relativistic jets \cite{Ando2005}. Other scenarios consider the spin down of rapidly rotating newborn pulsars producing large electromagnetic fields, a model that can explain also the highest energy CR \cite{Fang2013}. The fact that the most promising SNe types (e.g. IIn, Ib/c or hypernovae) are rare and transient by nature puts them within reach of IceCube.

\subsection{Steady sources}

\begin{figure}
	\centering
\includegraphics[scale=0.8]{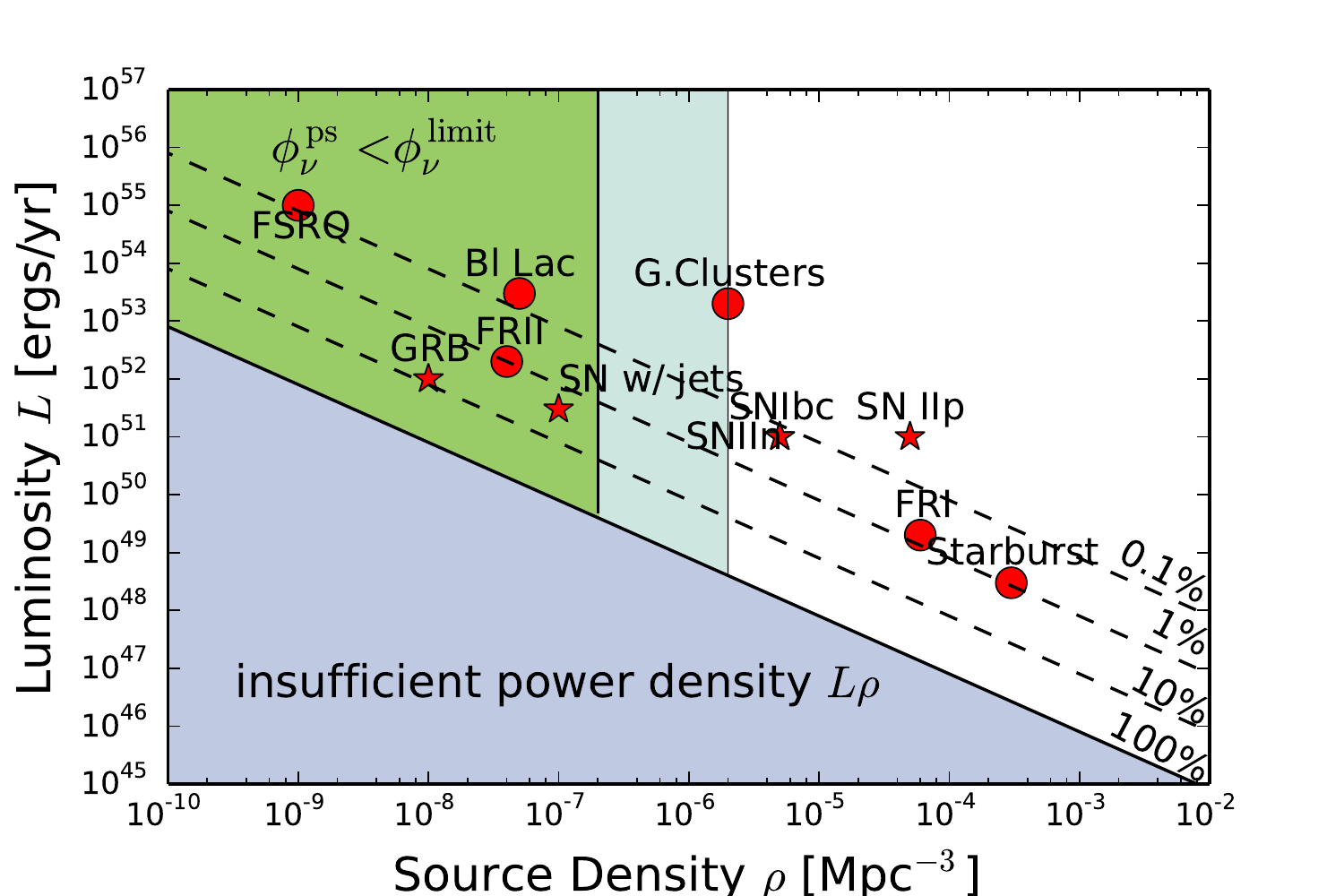}
\caption{
 Figure \ref{fig1} with  additional constraints on steady point sources added.  The figure shows the additional sources: Flat Spectrum Radio Quasars, Bl Lacs,  Fanaroff-Riley II galaxies, Galaxy Clusters,  Fanaroff-Riley I and Starburst galaxies (see caption of Fig.~\ref{fig1} for more explanation). }
\label{fig2}
\end{figure}

%{\bf Beschreibung der steady source sensitivity soll hier erfolgen...}
The previous considerations can be generalised to steady sources by dropping the time constraint of the multiplet search while increasing the required neutrino multiplicity to suppress the background of atmospheric neutrinos. With an angular resolution of $\sim\!1^\circ$ for muon neutrinos and within 3 years of observations with IceCube, the detection of $\sim\!$20 signal neutrinos from one source would be significant at the 3 $\sigma$ level. Stacked searches increase the sensitivity by a factor of a few, allowing to lower the number of signal neutrinos per source. The limit derived from the non-detection of $\geqslant10$ signal neutrinos from the same source yields a lower bound of $\rho\geqslant2\cdot 10^{-7}\text{Mpc}^{-3}$ as indicated in Fig.\ \ref{fig2}. 

Figure \ref{fig2} summarises the various astrophysical source populations that can potentially be responsible for the observed neutrinos, now also including steady sources. The sources populations are arranged according to their source density and the typical luminosity in electromagnetic radiation. The rarest and most luminous sources (in gamma rays) are Blazars (FSRQ and BL Lac), Fanaroff-Riley II (FR II) galaxies as well as GRBs. While all of these sources are candidates to produce the observed high-energy neutrinos, their number density is sufficiently low that one would have expected to see individual sources shining bright in the neutrino sky. The brightest sources within the sample would violate limits on (stacked) neutrino point sources, as well as limits on transient sources unless the population's density is larger than $\sim 10^{-7} {\rm Mpc}^{-3}$.
More frequent sources, such as Fanaroff-Riley I galaxies (FRI) or Starburst galaxies are not yet restricted by point source limits, since the total flux is produced by sufficiently many sources. 

\section{Summary and Outlook}
The evidence for high-energy astrophysical neutrinos has increased significantly since their discovery in 2013.  Due to their distribution in the sky, the observed neutrino events appear consistent with being of extragalactic origin. While individual sources have not yet been identified, even the absence of a steady point source and transient source signal  leads to severe constraints on the possible nature of the source.  The source density should be $\lesssim 10^{-7} {\rm Mpc}^{-3}$ for steady sources and $\lesssim 10^{-8} {\rm Mpc^{-3} yr^{-1}}$ for short transient sources (burst duration $< 100$~s), or individual bright sources would have been observed already. 

Currently,  studies are ongoing to better constrain the spectrum and flavour composition. About $1/3$ of the astrophysical neutrino flux should consist of $\nu_\tau$ and their identification  should eventually be feasible based on their distinct ``double-bang" signature \cite{db} (the distance between hadronic interaction vertex and tau decay is $\sim 50~ {\rm  m} \times (E_{\tau}/ \rm PeV)$ and hence might be resolvable for the highest energy events). Soon,  also other experiments hopefully reach the sensitivity to confirm IceCubes observation of astrophysical neutrinos.

Resolving the sources of high-energy neutrinos will be of particular focus for future studies. With  existing neutrino detectors alone, the anticipated gain in sensitivity appears not sufficient to firmly resolve the sources. However, through combining neutrino information with observations from radio, optical, X-rays to $\gamma-$rays, there is justified hope to significantly narrow down the possibilities, and eventually identify the source. %A new generation of more sensitive neutrino detectors covering the full sky will  allow to fully exploit the scientific potential of neutrino astronomy.

{}
\end{document}